        \documentclass[12pt]{revtex4} 

        \expandafter\let\csname equation*\endcsname\relax
        \expandafter\let\csname endequation*\endcsname\relax
        \usepackage{amsmath,bm}
        \usepackage{amsfonts} 
        \usepackage{mnsymbol}  
        \usepackage{graphicx}
        \usepackage{hyperref}
        \usepackage{url}
        \usepackage{lineno}  
        \usepackage{color}
        \usepackage{comment} 
        \usepackage{epstopdf}  
        
        \newcommand{\C}{\mathbb{C}}
        \newcommand{\eq}[1]{\begin{equation}#1\end{equation}}
        \newcommand{\suq}[1]{SU$_q$(#1)}

        \newcommand{\q}{$q$}
        \newcommand{\ra}{\rangle}

        \newcommand{\qdfmd}{$q$-deformed }

\begin{document}
    
    \title{Entanglement entropy in quasi-symmetric multi-qubit states}
    
\author{Zhi-Hua Li  and An-Min Wang }
\affiliation{Department of Modern Physics, University of Science and Technology of China, Hefei 230026, China}%
\begin{abstract}
        We generalize the symmetric multi-qubit states to their $q$-analogs, whose basis vectors are
        identified with the $q$-Dicke states.  We study the entanglement entropy in these states and
        find that entanglement is extruded towards certain regions of the system due to the
        inhomogeneity aroused by $q$-deformation.  We also calculate entanglement entropy in ground
        states of a related $q$-deformed Lipkin-Meshkov-Glick model and show that the singularities
        of entanglement can correctly signify the quantum phase transition points for different
        strengths of $q$-deformation.
    \end{abstract} 



\maketitle


\section{Introduction}

    
    Quantum entanglement as the resource of quantum
    information process is the genuine characteristic of quantum systems \cite{nielsen2010quantum}. 
    In the last decade, there have been enormous interests in studying entanglement of  
    quantum many body systems.
    The Hilbert space $(\C^2)^{\otimes N}$, taking $N$-qubit systems for example, grows
    exponentially large with $N$, so measuring of entanglement of arbitrary states in $(\C^2)^{\otimes
    N}$ is generally hard \cite{dur2000three}.
    One simpler task is to restrict to some particular subsets of $(\C^2)^{\otimes
    N}$, in which the computation of entanglement should be feasible. An example is the symmetric
    subspace, which is spanned by the states obtained by superposition of distinct permutations of
    state like $|11\cdots1 0\cdots0\ra$ with $k$ spin-up's and $(N-k)$ spin-down's.
    Pairwise entanglement and entanglement entropy have been obtained in these states \cite{wang2002pairwise,stockton2003characterizing}.
    Since the early works, the symmetric states have become a testing ground for new
    entanglement measures \cite{wang2003spin, wei2003geometric, devi2007characterizing} and for
    elucidating some features of entanglement structure in many-particle systems
    \cite{dur2000three, bastin2009operational,aulbach2010maximally}.
    The computation of entanglement in these states can also be used in the studies of quantum spin chain
    models whose Hamiltonian has the permutation invariance, an example being the
    Lipkin-Meshkov-Glick (LMG) model \cite{lipkin1965validity}. 
    In regard to the seminal works on the intriguing relations between entanglement and quantum phase transition
    \cite{osborne2002entanglement, osterloh2002scaling, gvidal2003entanglement,
    jin2004quantum,calabrese2004entanglement} (see \cite{amico2008entanglement} for a review),  the relation of anomaly of ground state quantum entanglement with
    quantum phase transition in the LMG model have also been established
    \cite{latorre2005entanglement,vidal2004entanglement,vidal2003entanglement,vidal2007entanglement,wichterich2010universality,zhang2013singularities}.
   
    Given that the symmetric states are amenable to the characterization of entanglement properties of quantum
    many-body systems and that they can be realized in experiments 
    \cite{prevedel2009experimental,wieczorek2009experimental}, it is theoretically desirable to extend these
    entanglement-computable states to a wider range for further analysis. 
    In physics, one particular way to extend certain physical models or states is the 
    quantum deformation through  quantum groups \cite{drinfeld1987proceedings,jimbo1985aq,kassel1995quantum}.
    Recently there have been some interests in studying entanglement in \q-deformed many body
    systems, such as the \q-deformed coherent
    states\cite{berrada2011entanglement,berrada2012bipartite}, \q-VBS states
    \cite{santos2012entanglement,santos2012entanglement_jpa}, \q-Werner
    states\cite{liu2013quantum} and some related models. The \q-deformed states usually exhibits
    some non-classical properties which provide new perspectives on the understanding of
    entanglement in many body systems and may have potential applications in certain quantum
    information tasks.  
    
    In this work we extend the symmetric states to a
    wider class of states depending on a deformation parameter $q$, whose basis is later identified
    with the $q$-Dicke states through the \suq2 quantum group. 
    The deformation breaks the permutation invariance in the original symmetric states, but 
    we show that the requirement of permutation invariance can be
    actually to some extent relaxed in the calculation of entanglement entropy. 
    Besides, these states don't  even preserve the cyclic or reflection invariance. As a result, we
    find that  entanglement is extruded to certain regions of the system.  This behaviour differs from the
    usual many body states studied in the literature where the entropy are mostly found to be
    symmetric about the middle bipartition.
    In addition we also study a corresponding \q-deformed LMG model. Surprisingly, the singularity in the
    entanglement entropy can still correctly signify quantum phase transition even in the presence of
    \q-deformation. 
        
\section{\q-analog of symmetric states}   \label{sec:define} 
    
        In order to define the \q-analog of symmetric states, let us recall the notations of 
    \q-analogs of some usual mathematical objects:  The \q-factorial is 
       \begin{equation}
            \label{eq:q:factorial}
            [n]! = [n][n-1]\cdots[1],
        \end{equation}
    where 
         \begin{equation}
               [x]\equiv \frac{q^x-q^{-x}}{q-q^{-1}},
               \label{eq:q:num}
               \end{equation}
    denotes the $q$-numbers and $q>0$. $[x]$ is invariant when change $q\leftrightarrow q^{-1}$.
    Then the $q$-binomial coefficients are
        \eq{ {n \brack m}= \frac{{[n]!}}{{[m]![n - m]!}}}
    In the limit of $q\to 1$, the above definitions all return to the normal ones: $[x]\to x$,
    $[n]!\to n!$ and $ {n\brack m} \to{n \choose m}$.
    
    Now we consider the following classes of states $|N,k\ra_q\in (\C^2)^{\otimes N},\,\,\,k=0,1,\cdots N$, which are
    extensions of the basis of symmetric states for each $q>0$ 
        \eq{
         \label{eq:q:symm:state} |N,k \ra_q \equiv 1/\sqrt{N \brack k} \;\;\sum\limits_{1 \le
        {i_1} < \cdots < {i_k} \le N} {
        {{q^{-k(N - k)/2 + \sum\nolimits_{l = 1}^{l = k} {{i_l}} {\rm{ - }}(k + 1)k/2}}}
        |0\cdots 0\mathop 1\limits^{\mathop  \vee \limits^{{i_1}} } 0 \cdots
        0\mathop 1\limits^{\mathop  \vee \limits^{{i_2}} } 0 \cdots 0\mathop 1\limits^{\mathop
        \vee \limits^{{i_k}} }0\cdots 0 \rangle } } 
    They include basis of the symmetric states for $q=1$.
    Largely speaking each state $|N,k\ra_q$ is a superposition of all distinct
    permutations of the state $|1\cdots1_k 0\cdots0_N\ra$ with nontrivial
    coefficients which are $q$'s powers (Let alone the overall normalization factor
    $1/\sqrt{N\brack k}$).  
    The powers of the $q$-factors for each permutations are seen being defined in the following rules:
    1) Starting from the ``word'' $11\cdots100\cdots0$ with k 1's followed by $(N-k)$ 0's, 
    its \q-factor is set $q^{-k(N-k)/2}$; 2) Each time a 1 crosses over a 0 from left to right will produce a
    factor $q$. 
    From these, as for a generic word having 1's at positions $i_1,i_2,\cdots,i_k$, the power on $q$ is 
    $ -{k(N-k)/2} + \sum\nolimits_{l = 1}^{l = k} {({i_l} - l)}$, besides, the squares of these
    \q-factors sum up to $1/{N\brack k}$, thus finally yielding the properly normalized form of
    eq.\eqref{eq:q:symm:state}.
    One may also put the rules simply: The more 1's distributed on the rightmost positions in the
    word, the larger the power on $q$ \cite{note1}.  
    These rules are most easily anticipated by looking at a simple example, e.g.,
    \eq{ |4,1\ra_q= {1}/{\sqrt {[4]} }({q^{-3/2}}|{1000}\rangle  + {q^{-1/2}}|0100\rangle  + {q^{ 1/2}}|0010\rangle  + {q^{ 3/2}}|0001\rangle ),}
    with $[4] = {q^{-3}} + {q^{-1}} + {q^{1}} + {q^{3}}$. 
    The \q-factors in essence keep  track of how the 1's and 0's are permuted 
    and can be regarded as inhomogeneous weights for each permuted words.
    As a result, each states $|N,k\ra_q$ are no longer permutation invariant, thus we may also call them and
    \emph{all of their superpositions} the quasi-symmetric states. 
    Additionally, we stress that the \q-factors even break in special the
     cyclic subgroup and reflection subgroup of the symmetric group. This inhomogeneity will cause unusual
     behaviours in bipartite entanglement. 
    
        The above definition of quasi-symmetric states is combinatorics flavored. 
    Actually they are identified with the \q-Dicke states within the \suq2 quantum group
    formalisms.  The algebraic approach will facilitate later the calculation of the entanglement property of the
    states and the definition of related physical models. 
    %
    %
    
    Recall the definition of the \suq2 algebra, which is generated by
    the operatoers $S^+$, $S^-$ and $S^z$ subject to relations:
    \begin{equation} \label{eq:suq2}
    \begin{split}
            [{S^z},{S^ \pm }] &= {S^ \pm }\\
            [{S^ + },{S^ - }] &= [{2S_z}] \equiv  \frac{{{q^{{2S^z}}} - {q^{ - {2S^z}}}}}{{q - {q^{
            - 1}}}},
    \end{split}
    \end{equation}
     Its representations are in parallel with SU(2) algebra\cite{jimbo1985aq}: Each irreducible
     representation is labeled by the total $q$-spin $S \in \{0,\frac{1}{2},1,\frac{3}{2},\cdots\}$.
     For a fixed $S$, the representation space $V^S$ is of dimension $2S+1$ and spanned by the
     so called \q-Dicke states $\{|S,M\rrangle_q, M=-S,-S+1,\cdots,S\}$ (Here using $|\cdot\rrangle$ to avoid
     confusion with the notation in eq.\eqref{eq:q:symm:state}). The representation matrices of
     the generators are given through:
        \begin{equation}   \label{eq:suq2:generator:mat}
        \begin{split}
            {{S}^{z}}|S,M\rrangle_{q} &= M|S,M\rrangle_{q} \\
            {{S}^{\pm }}|S,M\rrangle_{q} &=\sqrt{[S\mp M][S\pm M+1]}|S,M\pm 1\rrangle_{q}
        \end{split}
        \end{equation}
    In particular, the fundamental representation for the \suq2 algebra is trivially identical to that
    of  the SU(2) algebra: The representation space is $V^\frac{1}{2}\equiv\{|1\ra, |0\ra\}$
    ($|1\ra$ and $|0\ra$ are spin up and down respectively), and the representation matrices are
     $s^z\equiv \frac{1}{2}\sigma^z, s^\pm\equiv\sigma^x\pm i\sigma^y$, with $\sigma^\alpha,
     \alpha={x,y,z}$ being Pauli
    matrices. 
    
    A well known fact about SU(2) Lie algebra is that, all its irreducible representations can be
    constructed by coupling of several spin-1/2 representation matrices. Similar situation also
    holds for the \suq2 algebra. This fact entails connection between formalisms of Lie algebras or quantum algebras
    with the multi-qubit systems, so we restate this fact in detail. The only subtlety in the \suq2
    case is that in order that the coupled \q-spins to respect the \suq2 relations \eqref{eq:suq2},
    one must utilize the comultiplication 
    %
    %
    %
    $\Delta: SU_q(2)\to SU_q(2)^{\otimes 2}$,
    \begin{align}
          \Delta ({S^z}) &= {S^z} \otimes 1 + 1 \otimes {S^z}  \nonumber \\
          \Delta ({S^ \pm }) &= {q^{{S^z}}} \otimes {S^ \pm } + {S^ \pm } \otimes {q^{ - {S^z}}}
     \end{align}
     Then the coupling of $N$ \q-spin-1/2 operators are obtained by  acting $\Delta$ on $s^{\pm,z}$ 
     for $N$ times: 
     \begin{equation}  \label{eq:comultiplication:N}
         \begin{split}
            {\tilde S^z}   &= \Delta^{N} (s^z)    = \sum\limits_{i=1}^N {1 \otimes \cdots  \otimes
            1\otimes \mathop {s^z}  \limits^{\mathop  \vee \limits^i } \otimes 1 \otimes \cdots  \otimes 1} \\
            {\tilde S^\pm } &=\Delta^{N} (s^\pm)  = \sum\limits_{i=1}^N { {q^{{s_z}}} \otimes
            \cdots  \otimes {q^{{s_z}}}\otimes\mathop {{\sigma ^ \pm }}\limits^{\mathop  \vee
            \limits^i } \otimes {q^{
            - {s_z}}} \otimes  \cdots  \otimes {q^{ - {s_z}}}}  
        \end{split}
    \end{equation}
    so that $\tilde S^{\pm,z}$ are some particular representation matrices for $S^{\pm,z}$. Next is
    to determine which representation the operators $\tilde S^{\pm,z}$ belong to: 
    First note the state $|N,0\ra_q = |0_10_2\cdots 0_N\ra$ is the lowest weight vector for this
    representation as $\tilde S^-|N,0\ra_q=0$.
    Successive action on
    $|N,0\ra_q$ by $\tilde S^+$ for $k$ times $k=0,1,2\cdots$ will generate a whole basis for this
    representation, which can be calculated directly by using
    \eq{    \label{eq:spk} 
        \begin{array}{l}
        {\tilde S^{ + k}} = [k]!\sum\limits_{1 \le {i_1} < {i_2} <  \cdots  < {i_k} \le N} {{q^{k{s_z}}} \otimes  \cdots  \otimes {q^{k{s_z}}} \otimes \mathop {{\sigma ^ + }}\limits^{\mathop  \vee \limits^{{i_1}} }  \otimes {q^{(k - 2){s_z}}} \otimes  \cdots  \otimes {q^{(k - 2){s_z}}}} \\
        \;\;\;{\kern 1pt} \;\;\;{\kern 1pt} \;\;{\kern 1pt}  \otimes \mathop {{\sigma ^ + }}\limits^{\mathop  \vee \limits^{{i_2}} }  \otimes {q^{(k - 4){s_z}}} \otimes  \cdots  \otimes \mathop {{\sigma ^ + }}\limits^{\mathop  \vee \limits^{{i_k}} }  \otimes {q^{ - k{s_z}}} \otimes  \cdots  \otimes {q^{ - k{s_z}}}
        \end{array}}
    The result of the calculation turns out to be that $\tilde S^{+k}|N,0\ra_q$ is nothing but expressions of the
    quasi-symmetric states $|N,k\ra_q$ (up to a normalization factor) for $k\leq N$ and equals 0
    when $k>N$. 
    So the dimension of this representation  is  $N+1$,  and considering the uniqueness of the
    irreducible representation of \suq2 \cite{jimbo1985aq}, this determines the representation 
    to be $S=N/2$. 
    Finally, according to \eqref{eq:suq2:generator:mat} we conclude the identification of the quasi-symmetric states
    with the \q-Dicke states $|N,k\ra_q \equiv |S=N/2,M=k+N/2\rrangle_q $.
    With this identification, we can utilize the \suq2 algebra to calculate entanglement properties
    of the quasi-symmetric states and construct models that generate these states, as shown in
    following sections.   
    
    %
\section{bipartite entanglement of quasi-symmetric states}  \label{sec:state}
    
    A generic $N$ qubit system can be split in a variety of ways. 
    In this paper we only consider bipartition of the system to $A$ and $B$ subsystems: 
    $A=\{1,2,\cdots,L\}$ and $B=\{L+1,\cdots,N\}$ for $L=0,1,\cdots N$.
   Then to compute the entanglement entropy ($EE$) of the states
   $|\psi\ra\in (\C^2)^{\otimes N}$, one way is to get the Schmidt decomposition:
      \eq{ 
       |\psi\ra = \sum\limits_i {{\lambda _i}|{\psi _{A,i}}\rangle  \otimes |{\psi _{B,i}}\rangle } 
       }
    where ${\psi _{A,i}}\in (\C^2)^{\otimes L}$ and  ${\psi _{B,i}}\in (\C^2)^{\otimes (N-L)}$. These
    vectors coincide with the eigenvectors of the corresponding reduced density matrix ${\rho _{A/B}}
    = {\rm{t}}{{\rm{r}}_{B/A}}(|\psi \rangle ) = \sum\nolimits_i {\lambda _i^2|{\psi
    _{A/B,i}}\rangle \langle {\psi _{A/B,i}}|} $. 
    So the
    entanglement entropy is \eq{ \label{eq:EE}
      {S_{N,L}} = \sum\limits_i {{\lambda _i^2}\log_2({\lambda _i^2})} }

    In the computation of the bipartite entanglement of the symmetric states, 
    their Schmidt decomposition is mapped from the decomposition of the
    corresponding Dicke states \cite{unanyan2004bi,latorre2005entanglement}.
    The quasi-symmetric states can be computed in the same  approach, except for replacing
    everything with their \q-analogs: The decomposition of
    the \q-Dicke states $F:{V^S} \to {V^\mu } \otimes {V^{S - \mu }}$ is just the decomposition of
    \q-spin angular momentum, given by
        \eq{  \label{eq:qdicke:decomposition}
            |S,M{\rrangle _q} = \sum\limits_{\nu  =  - \mu }^\mu  {C_{\nu ,M - \nu ,M}^{\mu ,S - \mu
            ,S}|\mu ,\nu {\rrangle _q} \otimes |S - \mu ,M - \nu {\rrangle _q}}, }
   where $C$  is the \q-Clebsh-Golden coefficients and its explicit expression (written in the standard
   notation) is \cite{kirillov1991clebsch}            
        \begin{equation} 
            \begin{split}
             C^{j_1,j_2,j_1+j_2}_{m_1,m_2,m} 
             ={{\delta }_{{{m}_{1}}+{{m}_{2}},m}} & {{q}^{{{j}_{1}}{{m}_{2}}-{{j}_{2}}m_1}}  \\
             &\times\left\{\frac{[2{{j}_{1}}]![2{{j}_{2}}]![{{j}_{1}}+{{j}_{2}}+m]![{{j}_{1}}+{{j}_{2}}-m]!}{[2{{j}_{1}}+2{{j}_{2}}]![{{j}_{1}}-{{m}_{1}}]![{{j}_{1}}+{{m}_{1}}]![{{j}_{2}}-{{m}_{2}}]![{{j}_{2}}+{{m}_{2}}]!}\right\}^{1/2}
            \end{split}
         \end{equation}
   With the identification of \q-Dicke states for $S\to N/2$, $M\to k-N/2$, $\mu\to L/2$ and $\nu\to
   l-L/2$, eq.\eqref{eq:qdicke:decomposition} is mapped to the corresponding decomposition
   in the $N$ qubit system  
   $\tilde F:{(\C^2)}^{ \otimes N} \to (\C^2)^{ \otimes L} \otimes (\C^2)^{\otimes(N - L)}$, 
   given by
        \eq{
           |N,k{\rangle _q}
             =\sum\limits_{l = 0}^L {p_{N,k,q,L,l}^{1/2}|L,\;l{\rangle _q} \otimes |N - L,\;k - l{\rangle
             _q}},
            }
    where the Schmidt coefficients ${{p}^{1/2}_{N,k,q,L,l}}$ are translated from the \q-CG coefficients in
    eq.\eqref{eq:qdicke:decomposition}:
         \eq{
            {{p}_{N,k,q,L,l}}
             \equiv {C_{l - L/2,\;k - l - (N - L)/2,\; k-N/2}^{L/2,\;(N - L)/2,\;N/2}}
            ={{q}^{{kL-Nl}}} {N \brack l}{N-L \brack k-l}/ {N \brack k}
            }
    %
    %
    Note that ${{p}}$ is also the \q-hypergeometric distribution and it has the symmetries
   ${p_{N,k,q,L,l}} = {p_{N,N - k,1/q,L,L - l}} = {p_{N,k,1/q,N - L,k - l}}$. In the following we shall  write ${p_{q,l}}$
    as a shorthand for  ${p_{N,k,q,L,l}}$ when its meaning is clear from the context.
     \begin{figure}
          \centering
          \scalebox{0.55}[0.55]{\includegraphics{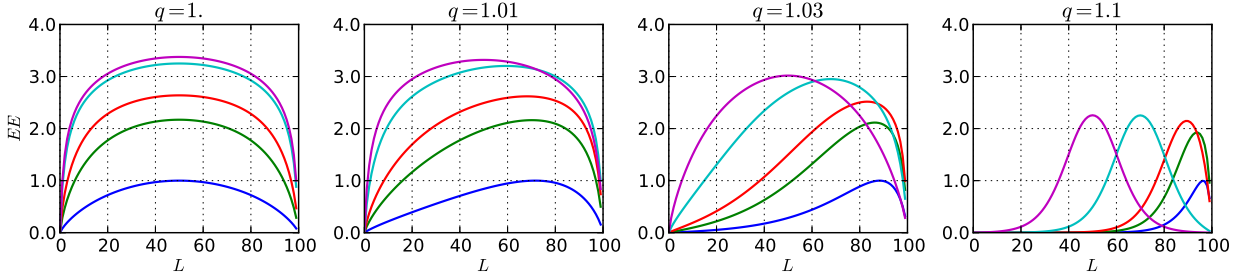}}
          \caption{\label{fig:EE:qdicke} Dependence of entanglement in $|N,k\ra_q$ on the position
          of bipartition $L$ for several values of $q$ and $k$: In each panel $N=100$ and (the left side of)
          the curves from bottom to top are $k=1,5,10,30,50$.}
          \end{figure}
    
     Once the Schmidt coefficients are obtained, the entropy is directly calculated through
     eq.\eqref{eq:EE}. 
     We numerically evaluate the summation in the entropy ${S_{N,k,q}}(L) =  -
     \sum\nolimits_{l = 0}^L {{p_{q,l}}{{\log }_2}{p_{q,l}}}$ for a system with 100 spins as shown in
     Fig.\ref{fig:EE:qdicke}.
     Note the symmetries in $p_{q,l}$ in turn leads to
     symmetries in $EE$: ${S_{N,k,1/q}}(L) = {S_{N,k,q}}(N - L)$ and ${S_{N,N - k,q}}(L) =
     {S_{N,k,q}}(N - L)$, so that, in Fig.\ref{fig:EE:qdicke}, we have let $L$ vary fully from $0$ to $N$ but select several
     values for $k$ and $q$ only in the ranges $k\leq N/2$ and $q\ge 1$, respectively. 
    Several properties can be observed: 1) For a fixed $N$ and $k$ \q-deformation won't increase the maximal values
    of entanglement that can be reached. And, as $q$ increase, the amount of entanglement is depressed in the whole
    system.  (Note in the limit $q\to\infty$, eq.\eqref{eq:q:symm:state} approaches a product
    state and entanglement vanishes); 2) The main effect of \q-deformation on entanglement is that,  
    as $q$ increases from 1, entanglement is extruded towards one side of the system. 
    And this effect is more obvious when $k$ deviates more from $N/2$. 
    The asymmmetry of the bipartite entanglement is mainly due to the inhomogeneity in
    $|N,k\ra_q$. 
    Nevertheless, note that changing $q\to q^{-1}$ amounts to  the reflection operation on $|N,k\ra_q$.
    So, in the figure the curves are extruded
    to the right, but if we change $q\to q^{-1}$ the curves would be just reflected about
    the line $L=50$ and thus towards the left side (This is also true if we change $k\to N-k$). 
    We have explored much larger values of $N$  and other combinations with $k$ and $q$ and
    find that these a few examples have already captured the essential entanglement properties in
    these classes of states.


    Having obtained the bipartite entanglement of each $|N,k\ra_q$, one may consider any of their
    superposition states:
    \eq{ \label{eq:psi}  
       |\psi \ra_q = \sum\limits_{k = 0}^N {{\alpha _k}{\rm{|}}N,k\ra_q } }
    For the bipartition of each $L$ as discussed above, the reduced density operator for $|\psi\ra_q$
    is readily written as
    \eq{ \label{eq:rho:L}        
    {\rho_L} = \sum\limits_{l,l' = 0}^L  {}
       {\sum\limits_{k = 0}^N {{\alpha _k}{{\bar
       \alpha}_{k-l+l'}}p_{q,l}^{1/2}p_{q,l'}^{1/2}|L,l\rangle \langle l',L|} },
        }
         which corresponds to a $(L+1)\times(L+1)$ matrix. So for $L\sim O(10^3)$, one may numerically
     diagonalize it to obtain its eigenvalues and then the entanglement entropy.
     Instead of studying entanglement entropy of some generic quasi-symmetric quantum states, we
     calculate entanglement of the ground states in a \qdfmd LMG model and emphasize relation between 
     entanglement and quantum phase transition even in presence of $q$-deformation.

\section{ground state entanglement in $q$-deformed LMG model}  \label{sec:qlmg}
     Avancini et. al. \cite{avancini1995phase} have proposed a $q$-deformed LMG model:
        
     \eq{  \label{eq:qlmg}
     H_q = \frac{h}{{4\sinh (\gamma /2)}}\sinh (2\gamma {\tilde S_0}) + \frac{\lambda}{2[N]}({\tilde
     S^{ + 2}} + {\tilde S^{ - 2}})}
     where $\tilde S^{\pm,z}$ are given by eq.\eqref{eq:comultiplication:N} and $\gamma = \ln q$.
     When $q=1$ it is the isotropic LMG model
     \eq{ H_1 =h\sum\limits_i {\sigma _i^z} + \frac{\lambda }{N}\sum\limits_{i < j} {\sigma _i^ + \sigma _j^{\rm{ + }}{\rm{ +
     }}\sigma _i^ - \sigma _j^ - }  
     }
     When $q\neq 1$, $H_q$ describe highly inhomogeneous systems, which are no longer permutation
     invariant nor even translation invariant.
     Nevertheless, since $H_q$ can be written in terms of the total \q-spin operators, it obviously has the \suq2 symmetry, i.e.  the
     Casimir element of \suq2 algebra 
     $C \equiv {S^ - }{S^ + } + {[{S^z} + \frac{1}{2}]^2} - {[\frac{1}{2}]^2}$ 
     commutes with $H_q$:
     \eq{ [H_q, C]=0, }
     So the total $q$-spin is conserved and $H_q$ is block diagonal in each total $q$-spin
     sector. It is known that the ground state belongs to the $S=N/2$ sector  spanned by
     $\{|N/2,-N/2\rrangle_q,|N/2,-N/2+1\rrangle_q,\cdots,|N/2,N/2\rrangle_q \}$. One may constraint $H_q$ in this
     sector and represent it by a $(N+1)\times (N+1)$ matrix under this basis. 
     Then, for systems with number of spins $N\sim O(10^3)$, which is large enough to extract the
     thermodynamic limit properties, one can directly diagonalize the
     Hamiltonian to get the exact ground state $|\psi\rangle_{gs}$.
         \begin{figure}
            \scalebox{0.75}[0.75]{\includegraphics{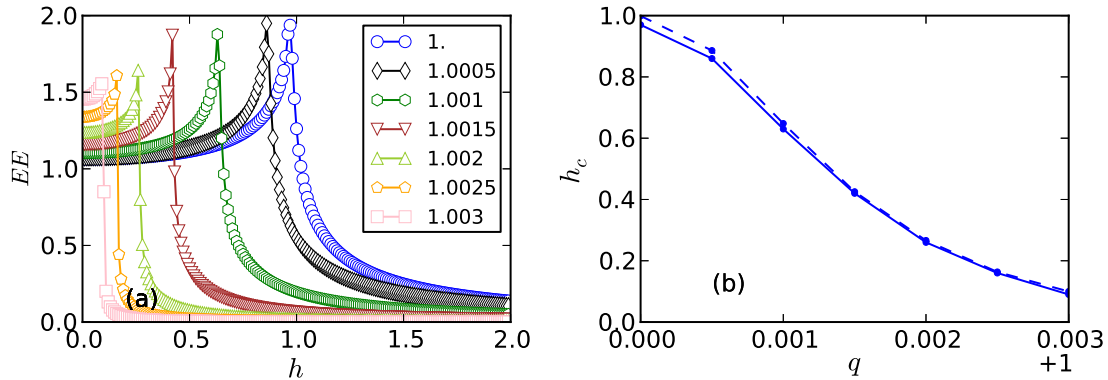}}
            \caption{ \label{fig:EE-qlmg}(a) Dependence of bipartite entanglement
            $S_{N,L}(|\psi_{gs}\ra)$ on magnetic field $h$ for different values of \q, with
            $N=1000,\,L=500$. (b) Critical magnetic field $h_c$ as a
            function of $q$: Solid line is determined through entanglement calculation; dashed through
            variation method.}
        \end{figure}
     
        Since the ground state is a generic quasi-symmetric state, the bipartite
    entanglement can be calculated through eq.\eqref{eq:psi}, \eqref{eq:rho:L} and \eqref{eq:EE}. We
    numerically calculate the ground state and its entanglement in this approach. 
    Without loss of generality, we fix $\lambda=1$ and let $H_q$ varies with $h$ and $q$.
    Bipartite entanglement  with $N=1000, L=500$ for $|\psi\ra_{gs}$ as a function of $h$ for different $q$ is shown in Fig.
    \ref{fig:EE-qlmg}(a). One can see that, for each $q$, the entanglement has a cusp at some value
    of $h$, which we denote by $h_c$. We have tested other bipartition by adjusting $L$ and find the results are
    qualitatively similar.
    With respect to previous studies of entanglement in LMG model \cite{latorre2005entanglement},
    one may expect that each cusp at $h_c$ indicates a quantum phase transition
    point. Note that $h_c$ decreases as $q$ increases.  
     This behaviour is consistent with the result of \cite{avancini1995phase} that increase $q$ will
     depress the phase transition. 
     Furthermore in \cite{avancini1995phase} the dependence of 
     $h_c$ on $q$ has been obtained analytically through mean field variational studies, which can be written as 
     \eq{ \label{eq:hc_vs_q:avancini} 
     {h_c(q)} = \frac{{[N - 1]}}{{[N]}}\frac{{2[\frac{1}{2}]}}{{1 + 2\,\sinh {{ {[\frac{\gamma}{2}(N -
         1) ]} }^2}}} } 
    We  extract the dependence of $h_c$ on $q$ for several $q$ values from the above entanglement
    calculations and compare it with
    eq.\eqref{eq:hc_vs_q:avancini}, as shown in Fig.\ref{fig:EE-qlmg}(b). It is shown that these two
    rather different approaches match well. So these two results confirm each other and this is  
    a strong evidence that $h_c$ are indeed quantum phase transition points. So we conclude that the
    cusp of entanglement entropy signatures
    quantum phase transition in the \q-deformed model.
    
          %

     
\section{Summary}
     In summary we have studied the entanglement entropy in a class of quasi-symmetric multi-qubit
     quantum states.  The entanglement entropy is obtained mainly by mapping the \q-spin angular
     momentum decomposition of \q-Dicke states to the Schmidt decomposition of multi-qubit states
     and is largely in parallel with the undeformed case.  
     %
     The main effect of \q-deformation on entanglement is that it extrudes the entanglement to 
     certain regions of the system.  
     An implication of the result is that one could use the deformation parameter to modulate the
     distribution of entanglement in certain systems. 
     In future works, one may also study the corresponding LOCC families \cite{bastin2009operational}, geometric
     entanglement and the  Majorana representation \cite{aulbach2010maximally} of the
     quasi-symmetric states, which would all together help to gain better understanding of
     entanglement structures in symmetric states. 
      In addition we have studied entanglement entropy in a \q-deformed LMG model related to these
     states. This model is a rare example in contrast with the bulk of models have been studied so
     far, which is not translation invariant and highly nonlinear.  In this regard, it is remarkable
     that  the singularities of the entanglement entropy is still able to detect the quantum phase
     transition point.

\section*{Acknowledgements}
     This work is supported by National Natural Science Foundation of China under Grant 
     No. 11375168.

%
     
%
    
%
%

    \bibliographystyle{apsrev}

    \section*{References}
    \bibliography{q-lmg-bibliography}

\begin{thebibliography}{36}
\bibitem[{not()}]{note1}
 \bibinfo{note}{One may alternatively define \eqref{eq:q:symm:state}
  such that the power on $q$ is higher if more 1's are distributed in the
  leftmost positions or simply change $q\to q^{-1}$.}
\end{thebibliography}

 

\end{document}